\documentclass[journal]{IEEEtran}
\usepackage{siunitx}
\usepackage[pdftex]{graphicx}
\usepackage{array}
\begin{document}

\title{A Novel Approach on Dielectric Barrier Discharge using Printed Circuit Boards}

\author{Jonathan~Gail,~Alisa~Schmidt,~Markus~H.~Thoma
\thanks{All authors were with the First Physical Institute, Justus Liebig University, 35392 Giessen,
Germany, e-mail: Markus.H.Thoma@exp1.physik.uni-giessen.de}}
\maketitle

\begin{abstract}
Drug resistant bacteria, prions and nosocomial infections underline the need of more effective sterilizing technologies. The cold plasma technology is expected to bring a benefit in this context \cite{jeon2015}.
Six different plasma sources, based on printed circuit boards, were evaluated fourfold. This include measurements of the power consumption, the ignition behavior by an ICCD-camera and ozone formation by absorption spectroscopy at 254\,nm. To evaluate the biocidal effect, four bacterial test series were performed with Escherichia coli. The entirety of the tests analyze the plasma inactivation process from the input parameters to the desired biocidal effect.\\
The discharge current and time resolved ignition behaviors indicated a simultaneous formation of filaments at the beginning of the negative half-cycle. The dynamics of the ozone production showed a saturated exponential growth upon a maximum value of 435\,ppm. Additionally, the microbiological test series unveiled differences between the plasma source concepts. A total reduction rate of \(\mathbf{10^{-4}}\) within a minute was achievable. An air flow through slits within the plasma sources destabilized the plasma.\\
Minor changes of the electrode geometry changed all measured parameters. Hence, to develop a pathogen inactivating plasma source, these results recommend a comb-shaped electrode design, which is laminated on a dielectric.
\end{abstract}


\section{Introduction}
The amount of nosocomial infections \cite{Nosokomial} and the rising number of resistant bacteria due to the application of antibiotics \cite{Talebi-Bezmin-Abadi:2019wz}, emphasize the importance of preventive hygiene in the medical sector. Since there are pathogens which develop a resistance to alcohol based sterilants \cite{Alkohol} or need a longer exposure to heat than it is used in the daily business \cite{Goldberg09}, the technology has to take a step towards more effective sterilization methods. One promising focus of recent research is the plasma sterilization \cite{Morfill_2009, MRSA}.\\
Plasma consists of electrons, ions, neutral gas and several reactive molecules. When working with the plasma afterglow of a cold atmospheric plasma (CAP) the charged particles recombine before they can reach a surface by diffusion. Hence, long-living reactive oxygen and nitrogen species (RONS) are considered to cause the main inactivating effect \cite{MetaPlasma,Pavlovich_2014}.  Low temperature, high inactivation rate and absence of plasma resistant pathogens is only a selection of the many advantages of the plasma technology. Despite that and the current research, there is still a lack of plasma sterilization applications especially in the medical sector. The low temperature of CAP enables the treatment of thermolabile surfaces like endoscopes, pharmaceutical goods or dental instruments \cite{Denis_2012,PD}. Their reprocessing is very complicated and susceptible to failures, which lead to a high rate of contaminated devices \cite{RUTALA2019A62}. This emphasizes the urgent development of new applications.\\
In this article six new designs of CAP sources are considered to improve the inactivation capability and manufacturability. Those are based on dielectric barrier discharge, which uses a dielectric to limit the discharge current and to distribute the charges over the surface. A specific setup is the surface micro discharge, where both electrodes (one as a grid) are attached to the dielectric and the plasma is generated within the gaps of the grid \cite{Li2019}. The power consumption drops from volume-, surface- to coplanar surface-DBDs, which have coplanar electrodes inside of the dielectric \cite{CEPLANT2002}. Adapting the coplanar surface-DBD technology, the sources are based on a thin printed circuit boards (PCBs) with comb-shaped electrodes \cite{CEPLANT2011}. To measure the effects of the different plasma sources, they become fourfold characterized by their electrical power consumption, time resolved ignition behavior, ozone production and inactivation capability. Two of six sources use an air-flow of \SI{1}{m^3/min} to improve the distribution of the RONS.

\section{Experimental setup}
\begin{figure*}[t!]
\centering
\includegraphics[width=0.95\textwidth]{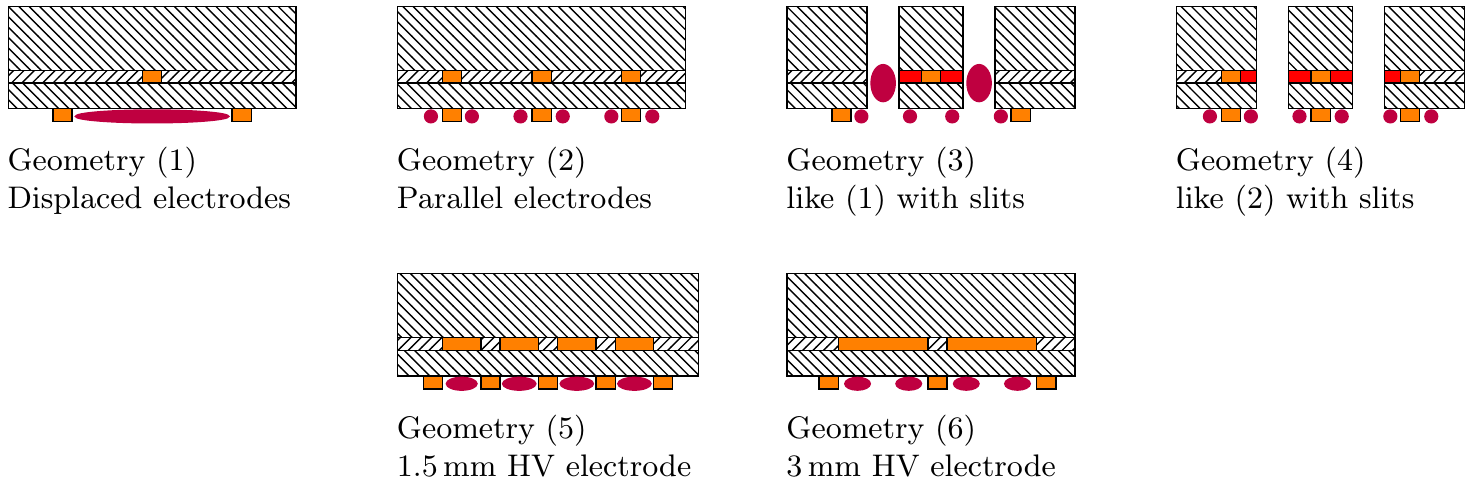}
\caption{Six different dielectric- and electrode geometries, violet: plasma, orange: copper electrode (lower electrode was grounded), black: FR-4, brown: epoxy resin, red: area of a potential breakdown, due to the slits, geometries (3) and (4) have spaces with only a small isolating epoxy part.}
\label{fig:Uebersicht}
\vspace*{4pt}
\hrulefill
\end{figure*}

The plasma sources were the result of a long development process, which was focused on a scalable and PCB-based design. As a substrate  the standard Flame Retardant FR-4 material, which is a dielectric made of an epoxy resin and a fiberglass inlay, is used. Copper layers are laminated on both sides, which can be etched in every desirable shape. Thus, the advantages are fast prototyping of different electrode geometries and the absence of air between the electrode and the dielectric. A drawback is that the FR-4 does not have a good ozone resistance. Additionally, the dielectric has a thickness of \SI{0.5}{mm} and a dielectric strength of \SI{20}{kV/mm}, which results in a maximum supply voltage of \SI{20}{kV} peak-to-peak (P2P)\cite{FR4}. Due to these conditions, the shelf life of these plasma sources was about 15\,h. Yet, the improvements of the inactivation capability and the reduction of the power consumption underline the relevance of the plasma research.\\
The voltage was supplied from a high voltage amplifier (Trek 10/40A-HS), which used the initial sine signal from a signal-generator (Rhode \& Schwarz HM8150) of \SI{3}{kHz}. The signal was amplified 1000-times up to a maximum value of \SI{10}{kV}.\\
Several tests during the development process, including the evaluation of the discharge behavior and antibacterial capability, lead to six different plasma sources, which are displayed in figure \ref{fig:Uebersicht}.

Since the diffusion of the reactive species to the contaminated surface is an inefficient process, an airflow has been proposed to have three advantages. First, the reactive species will reach the surface faster. Second, more reactive species will be directed towards the substrate and third, the plasma components with a short lifetime will have a contribution to the biocidal effect, because they reach the surface before dissociation.\\
The six plasma sources can be grouped in three conceptual geometry designs. The first has displaced electrodes, like (1 \& 3). The geometries (2) and (4) have parallel electrodes, which lead to a more uniform electric field and more homogeneous plasma. As figure \ref{fig:Uebersicht} shows, the area of the plasma glow is restricted to the edges of the electrodes. Combining the advantages of both geometries, the plasma concept of geometry (5 \& 6) has displaced electrodes without a horizontal distance between the electrodes. The high voltage electrode was wider than the grounded electrode. Hence, the electrons can accumulate over the dielectric and the plasma glow is larger than in case of (2 \& 4).

\subsection{Electrical Power Consumption}
The electrical measurements focus on the variation of the discharge current and its phase to the voltage. The oscilloscope (Tektronix DPO 2012B) acquired the voltage by the analog output of the high voltage amplifier and the current with a Rogowski coil, which enables the evaluation of the filaments \cite{Rogowski:1912tr}. From this point the analysis of the surface power density, power, reactive power, resistivity and capacitance was possible. The root mean square data of the voltage (RMS), current (RMS) and phase were averaged over 125\,000 cycles. Additionally, the surface temperatures of the plasma sources were measured to verify the positive dependency on the surface power density. Based on the findings of the electrical characterization, the operating voltage was chosen for later measurements. The operating frequency was constantly \SI{3}{kHz}. Since the high voltage amplifier has a power limit of \SI{400}{W} and the impedance was influenced by \(1/(\omega C)\), this frequency minimizes the need of the reactive power and thereby, enlarges the range of possible voltages, while the generated plasma was still homogeneously.

\subsection{Time Resolved Ignition Behavior}
Analyzing the ignition behavior of the filaments makes a very high resolution of the time necessary. This can be achieved by triggering and shifting the start of the frames. The photons were recorded by the ICCD-camera PI-MAX 4 from Princeton Instruments with an intensifying factor of 100, which was needed because of the weak plasma glow.\\
To evaluate 100 frames of one sine period with a frequency of \SI{3}{kHz}, it was necessary to have a gate width of \SI{3333}{ns}. In every period one frame was recorded. The gate delay was constantly increasing by \SI{3333}{ns} after each frame. Hence, every frame can be mapped to a defined position within one period, which was fully recorded after 100 frames. This enables the analysis of filaments and their formation.

\subsection{Ozone Production}
Besides the other reactive species, ozone was expected to have a major influence on the inactivation performance \cite{Shimizu2012,pavlovich2013ozone}. Therefore, the ozone concentration was measured for every plasma source at every possible operational voltage by absorption spectroscopy. This revealed the dependencies on power consumption and geometry. The measurement was made within a closed volume of \SI{5}{liters}.\\
The principle of the absorption spectroscopy uses UV-light with a wavelength of \SI{254}{nm}. A deuterium lamp (D-2000, Mikropack) generates a wide UV spectrum. The intensity at the wavelength of \SI{254}{nm} was measured by a spectrometer (OceanOptics HR-4000). This wavelength is in the middle of the Hartley-band of ozone, which has a high photo-absorption cross section of \(\sigma_{\lambda}=1.15\cdot10^{-21}\SI{}{m^2}\) and the other molecules do not interfere there \cite{ozone}. Hence, only this wavelength becomes absorbed by the ozone in the plasma afterglow. For this purpose, the experimental setup has an absorption path, with a length of \(l=\SI{20}{mm}\). This path was placed \SI{18}{mm} above the plasma source.
Within this defined length, the UV-light is send through the ozone containing plasma afterglow. At the end of the optical measurement setup, the intensity was monitored before the plasma ignition, during the operation of the plasma source and after the plasma was extinquished. This gives a typical U-shaped sequence, which enables the comparison of the initial reference intensity (\(I_0\)) and the intensity during plasma operation (\(I_1\)).\\
Based on the law of Beer-Lambert, this intensity ratio (\(I_0/I_1\)) can be correlated to the absorption events / ozone concentration (\(c\)).

\begin{equation}
\log_{10}(I_0/I_1) = \sigma_{\lambda}\cdot c\cdot l\label{eqn:BL}
\end{equation}

The plasma was ignited for minimum 15 minutes to evaluate the dependencies of the ozone dynamics on the heating and quenching effect. The quenching effect is the production of nitrogen species instead of ozone \cite{Shimizu2012}. After extinguishing the plasma, the ongoing measurement proofs if the reference intensity was reached after the dissociation of the ozone.\\
To achieve a good signal to noise ratio, an integration time of \SI{400}{ms} for each data point was used in every measurement. Also an electric dark correction was applied to reduce the offset of the sensor.

\subsection{Inactivation Capability}
Based on the measurements of the electrical characterization, the operational voltage was chosen respectively to the power surface density. The power density of the four sources (1) to (4) was set to \SI{0.427}{W/cm^2}, because with these settings, (2) produces the highest ozone concentration. Due to the restricted operational voltage of \SI{6}{kV} for geometry (6), the power density of the third plasma source concept (5 \& 6) was set to \SI{0.288}{W/cm^2}. For the evaluation of the inactivation performance, a micro-biological study on non-pathogenic strain K-12 of Escherichia coli (E. coli) was performed. To prepare the experiment, the bacteria were transferred from an agar plate to \SI{10}{ml} of the nutrient medium lysogeny broth (LB) via inoculation loop. After mixing on a test tube shaker (Genie Vortex 1), the solution was incubated at \SI{37}{\degree C} for \SI{24}{h}. Then, the optical density (OD\(_{600}\)) was set to 1 by diluting the culture with pure LB medium. The OD is the relative optical transmission of the bacteria culture to the pure medium at a wavelength of \SI{600}{nm}.\\
The resulting solution was diluted down by \(10^{-4}\) for the exposition to plasma and down by \(10^{-6}\) for the control measurement. Four series of tests were applied to every plasma source. The bacteria were spread over the agar plates with a Drigalski cell spreader. Each test series contained of 108 agar plates with 18 per plasma source. This number contains of 2 dilutions at 3 exposure times with 3 agar plates per case.\\
For the exposure, the plasma source was placed onto a frame \SI{2}{cm} above the petri dish. The resulting volume around the petri dish is \SI{0.2}{l}, where ozone saturation was reached after \SI{3}{s}. Hence, the effect of an increasing concentration during the exposition can be neglected.\\
Every colony forming unit (CFU) was counted manually after \SI{24}{h} and \SI{48}{h} of incubation. This enables the differentiation of fast growing colonies in the first step and identification of slow growing colonies in the second step. However, the relative difference of both measurements has never exceeded \SI{10}{\%}.

\section{Results}
\subsection{Electrical Power Consumption}
\begin{figure*}[t!]
   \centering
\begin{minipage}{0.4\textwidth}
   \centering
%   \text{Average of 4 cycles}
   \includegraphics[width=\textwidth]{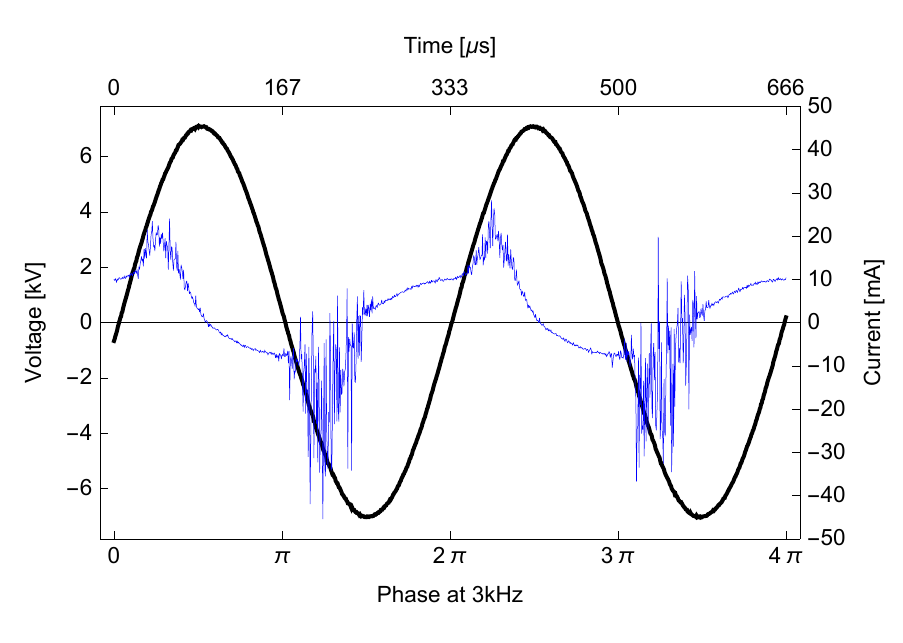}
\end{minipage}
\hspace{0.08\textwidth}
\begin{minipage}{0.4\textwidth}
   \centering
%   \text{Average of 32 cycles}
   \includegraphics[width=\textwidth]{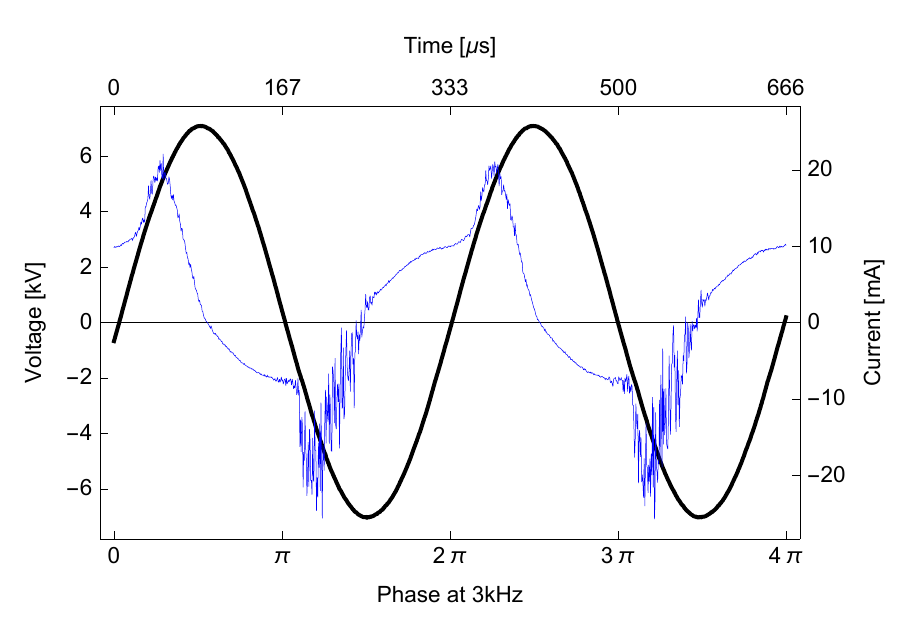}
\end{minipage}
\caption{Operational voltage (sine signal) and discharge current (signal with spikes) of plasma source (1) at \SI{3}{kHz} and \SI{14}{kV}; averaged over 4 (left) and 32 (right) cycles}
\label{fig:UI}
\vspace*{4pt}
\hrulefill
\end{figure*}

The operational voltage was set as a P2P-value at the sine generator. With the voltage, current and phase, the active power, capacitance and resistance were derived. A frequency of \SI{3}{kHz} enabled a homogenous plasma glow and reduced the reactive power, which limited the combinations of available voltages and frequencies. \\
Since the current was defined by the plasma source, only voltage and frequency were the input parameters. Therefore, the power was proportional to the square of the voltage, which can be verified for every plasma source.\\
The free electrons in the plasma influenced the capacity and resistance, which were evaluated. Additionally, the ratio of the active power to the surface gives the surface (active) power density, which can be chosen from \SI{0.05}{W/cm^2} to \SI{0.57}{W/cm^2} with respect to the plasma source concept. The surface power density was expected to be a predictor of the ozone concentration \cite{Shimizu2012,pavlovich2013ozone}.\\
Increasing supply voltages lead to a higher capacity and lower resistance, because the number of free electrons increased. Except in case of plasma source (5), which shows an increasing resist. This could be correlated to the curved filaments, which will be mentioned in the following. Due to the resistive heating, the temperature of the plasma sources reached up to \SI{64}{\degree C} with increasing supply voltages.\\
The shape of the measured discharge current reveals the dynamic of the filaments and was associated to the frames of the time resolved ignition analysis. Figure \ref{fig:UI} shows the operational voltage (sine) with \SI{14}{kV} P2P at \SI{3}{kHz} and the discharge current (signal with spikes). Its course was representative for all plasma sources and reveals two plasma ignitions during one period of the voltage. Increasing the voltage, causes stronger discharges, which unveil in higher currents. The spikes indicate the formation of filaments in the plasma \cite{LowTPlasma}. Even if the negative current has a stronger gradient and spikes, it was not possible to clarify, if it was caused by stronger and/or more filaments. Therefore, images of the time resolved ignition behavior with the ICCD-camera were made.
Due to the high variation of spikes, the data has to be averaged for the trigger. Figure \ref{fig:UI} compares two different averages of the same object and operational parameters. The strong smoothing effect of an average over 32 cycles was recognizable. This reveals that the spikes were statistically distributed and can be removed by averaging. The high gradient of the negative current emerges clearer with a stronger averaging. It was also visible, that the discharge starts close to the zero crossings.\\
The feed through led to higher discharge currents, especially during the ignition phase of the plasma. Since the air flow did not influence the discharge current at all, differences of the power consumption were a result of the geometries. Even though the visible inspection shows a destabilization of the plasma, which reduces the number of filaments, when the airflow was applied, this was not recognizable in the measured currents.

\subsection{Time Resolved Ignition Behavior}
\begin{figure}[t!]
   \centering
   \includegraphics[width=0.24\textwidth, height=6cm]{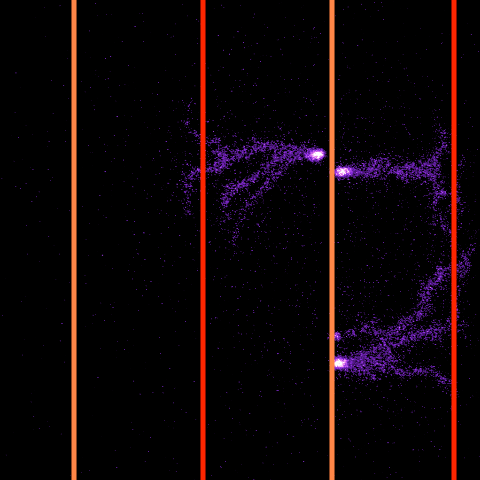}
   \includegraphics[width=0.24\textwidth, height=6cm]{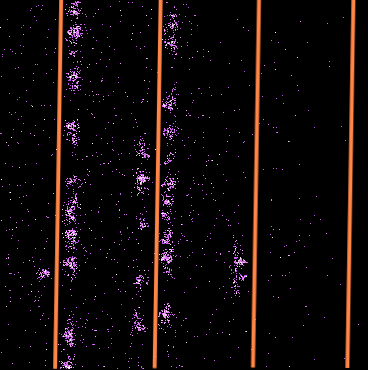}
   \caption{False color images of geometries (1) and (2) within the negative half-cycle, ground electrode in orange, high voltage electrode in red (left), grounded electrode is on top of the high-voltage electrode (right)}
\label{fig:ICCDreduziert1}
\end{figure}

\begin{figure}[t!]
   \centering
   \includegraphics[width=0.24\textwidth, height=6cm]{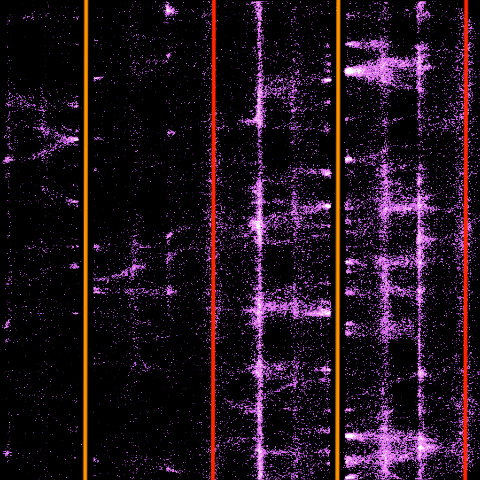}
   \includegraphics[width=0.24\textwidth, height=6cm]{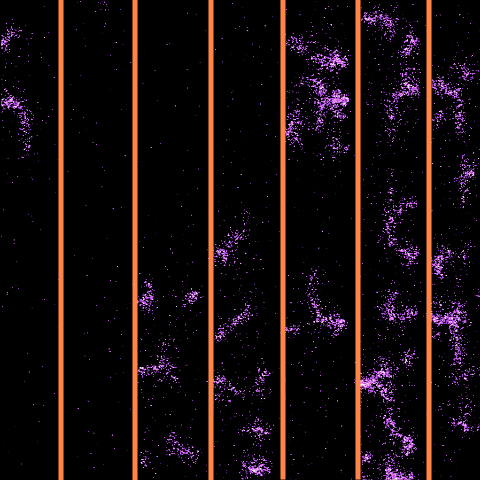}
   \caption{False color images of geometries (3) and (5) within the negative half-cycle, ground electrode in orange, high voltage electrode in red (left) and whole space between the ground electrodes (right)}
\label{fig:ICCDreduziert2}
\end{figure}

\begin{figure}[t!]
\centering
\begin{minipage}{0.49\textwidth}
   \centering
   \includegraphics[width=0.49\textwidth, height=6cm]{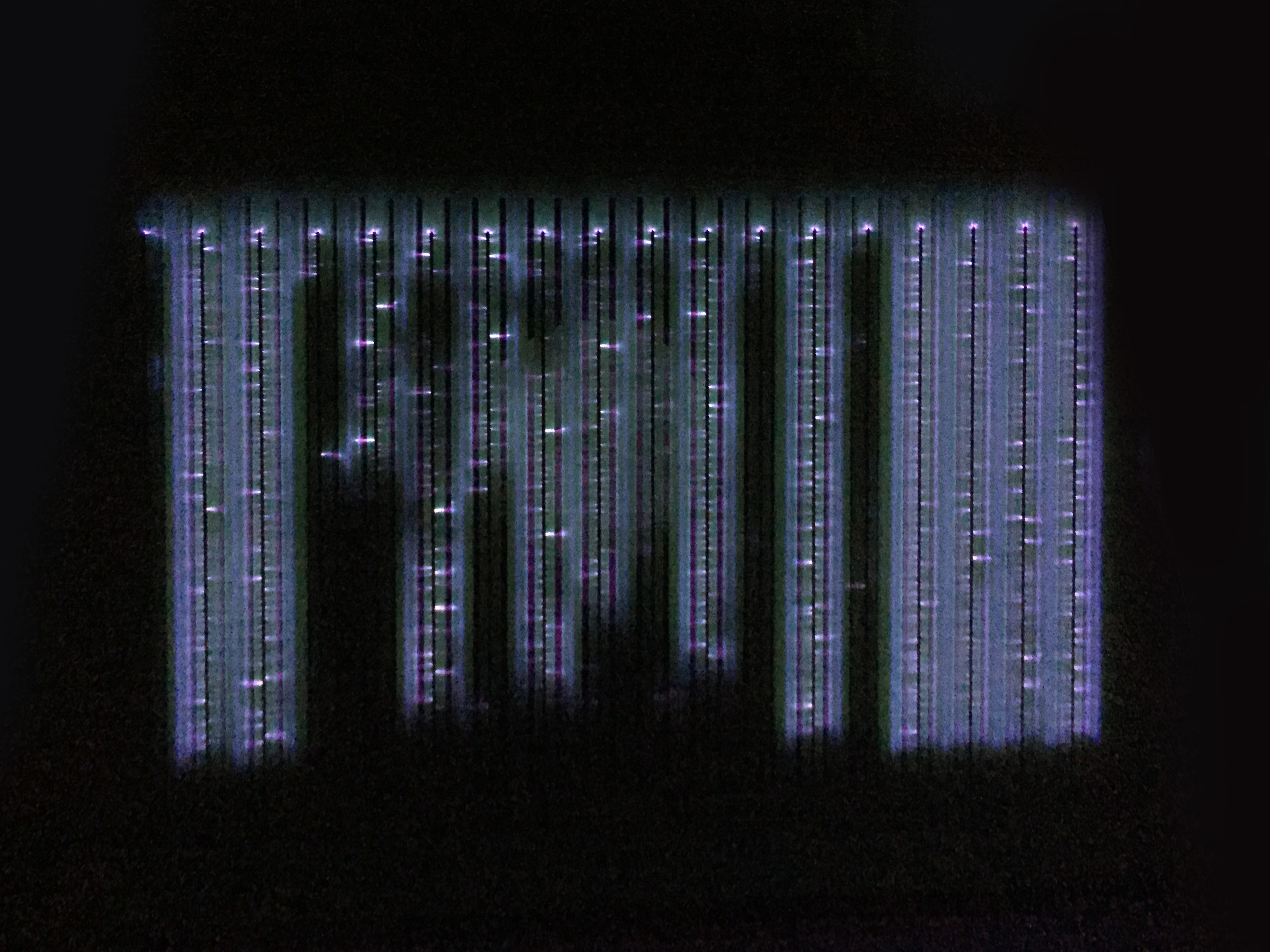}
   \includegraphics[width=0.49\textwidth, height=6cm]{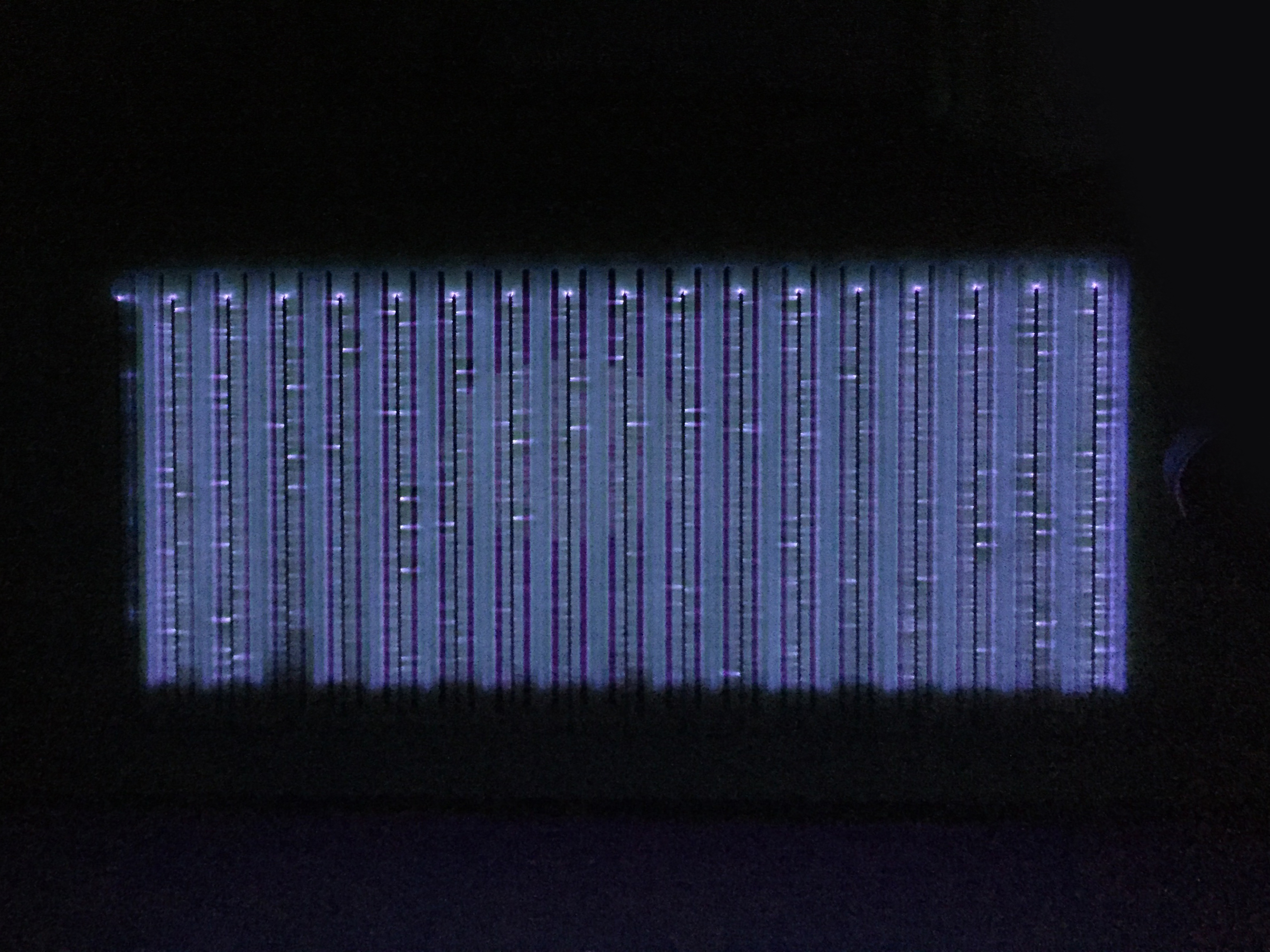}
\end{minipage}
\caption{Destabilizing effect at geometry (3) with (left) and without (right) the applied air flow}
\label{fig:Geo3Airflow}
\end{figure}

The results were 1024 \(\times\) 1024 matrices with an intensity value for each pixel. These data were plotted as false-color images. Black areas indicate no emissions from the plasma and violett to white areas represent the detection of photons. Those were caused by the emission of the plasma glow.\\
Figures \ref{fig:ICCDreduziert1} and \ref{fig:ICCDreduziert2} show the most relevant frames of the plasma sources in the negative half-cycle, which show the filaments and their systematical distinctions in dependency of the geometry. Other frames and results can be requested from the authors. Due to the distance between the electrodes, the plasma of geometry (1) forms arc-like filaments. In the left part of figure \ref{fig:ICCDreduziert1} several ionization events unite to one streamer towards the grounded electrode. Another frame (not shown here) shows the opposite direction of the electrons. The electrons accumulate above the dielectric covered high voltage electrode.\\
Figure \ref{fig:ICCDreduziert1} shows the simultaneous discharge of the filaments directly after the zero-crossing of the voltage. This clarifies, that the higher discharge current in the negative half-cycle was caused by more coincident filaments. This behavior can be verified for every plasma source. The residual charges on the dielectric surface were responsible for that, because they increased the electric field. Additionally, this caused filaments slightly before the zero-crossing.\\
Characteristic for concept (2) were the \SI{1}{mm} wide areas at the edges of the electrodes, where the plasma ignites. There were only few filaments, which reach up to the dark space between the small lines of the plasma glow.\\
One of the most interesting frames were displayed in the right part of figure \ref{fig:ICCDreduziert2} for geometry (3). The ignition behavior of this plasma source was remarkable. At first, the plasma ignites above the dielectric surface and higher voltages lead to a plasma formation within the slits in the dielectric. The mentioned frames unveil, that this behavior was a result of electron accumulation at the sides of the slits. These edges act like virtual electrodes. After increasing the voltage, the electrons reach to the opposite edge of the slits and form a plasma glow (displayed). The air flow through the slits lead to a destabilization of the plasma. Figure \ref{fig:Geo3Airflow} show the plasma source (3) with and without an applied airflow, which causes a weaker plasma and spaces without a discharge. It is proposed, that the streamer was deformed and extinguished.

Since the positions of the filaments were the same, the air flow does not affect the ignition behavior of geometry (4), because its filaments does not span over the slits. Based on the first two plasma sources, the electrode concept (5 \& 6) shows the desired results. Since the edges of the high voltage and grounded electrodes do not have a horizontal distance, the plasma ignites as homogeneously as plasma source (2). The \SI{1.5}{mm} wide dielectric covered electrodes spread the plasma uniformly over the entire dielectric surface. The length of the filaments in the right part of figure \ref{fig:ICCDreduziert2} indicate, that they were longer than the half of the high voltage electrode width. Increasing this width to \SI{3}{mm} lead to the formation of a darker area in the middle. The electrons were not fast enough to fill the entire surface with filaments.\\
Very interesting is the unique shape of these filaments. Figure \ref{fig:ICCDreduziert2} shows in the right frame curved paths of the streamers above the dielectric. This was caused by the structure of the electric field generated by these plasma sources.

\subsection{Ozone Production}
\begin{figure}[t!]
   \centering
   \includegraphics[width=0.49\textwidth]{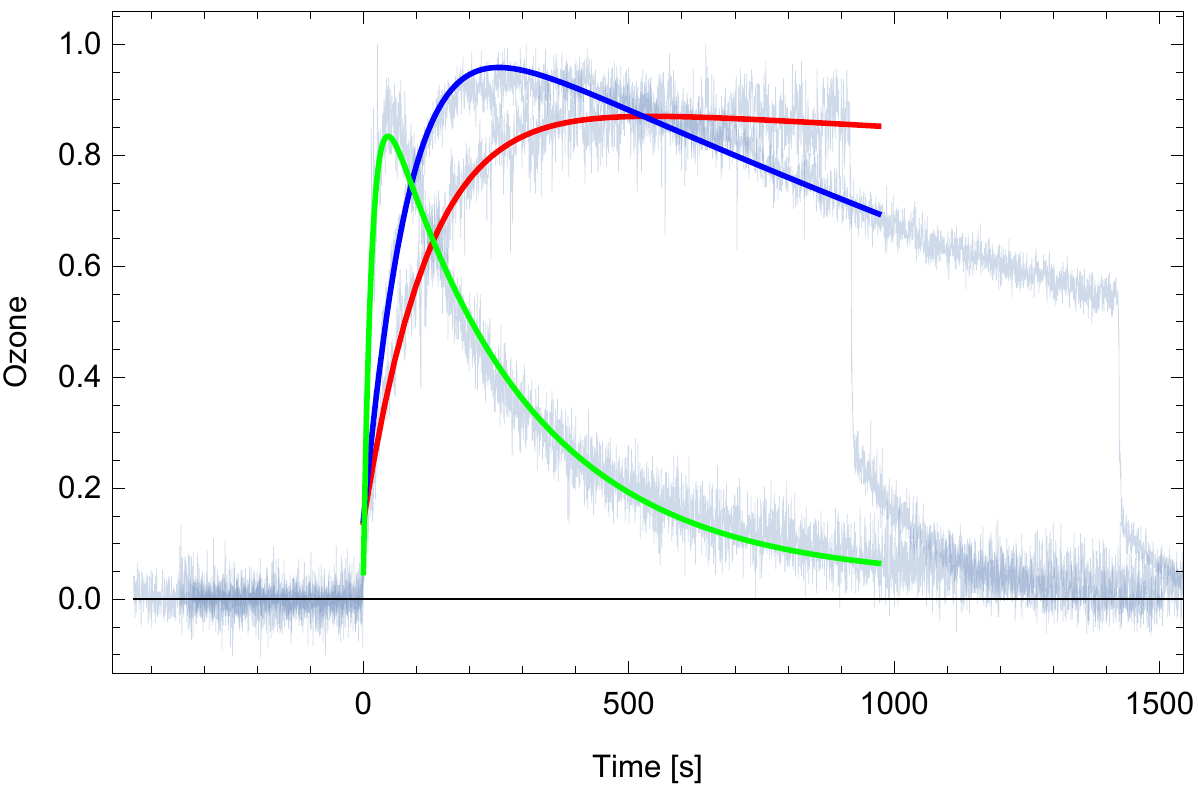}
   \caption{Characterisitic dynamics of the ozone production, Geometry (2) \SI{10}{kV} (red); Geometry (2) \SI{16}{kV} (blue); Geometry (5) \SI{12}{kV} (green)}
   \label{fig:Trends}
\end{figure}

With the results of the electrical power consumption, the ozone concentration was measured at all available operational voltages. During all measurements the reference intensity was reached, when the plasma was shut off, which is visible in figure \ref{fig:Trends}. There the black line indicates the reference concentration, based on the initial intensity (\(I_0\)) in the Beer-Lambert's law, see equation \ref{eqn:BL}.

\begin{figure*}[t!]
   \begin{minipage}{0.48\textwidth}
\includegraphics[width=0.8\textwidth]{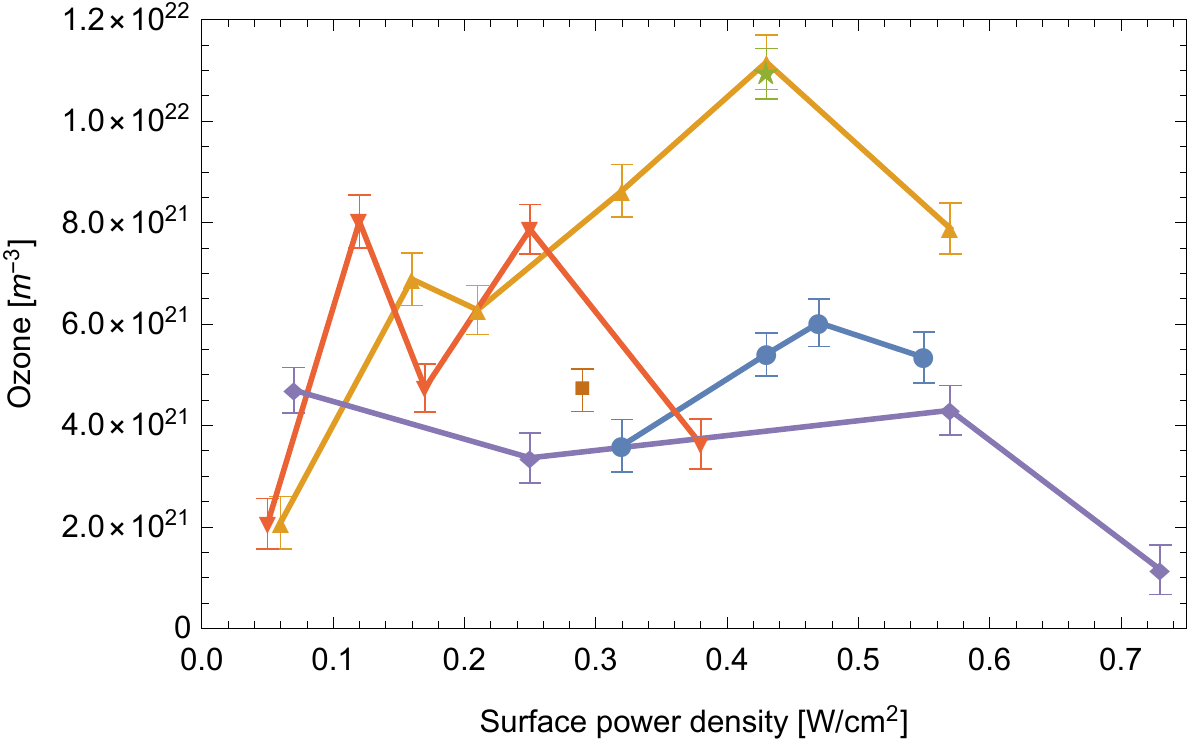}
   \begin{picture}(1,1)
	\centering
		\put(5,35){\includegraphics[width=1.2cm]{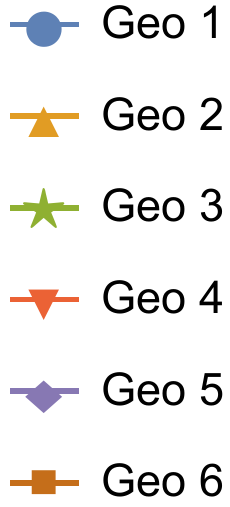}}
	\end{picture}
\end{minipage}
   \begin{minipage}{0.48\textwidth}
   \centering
   \includegraphics[width=0.8\textwidth]{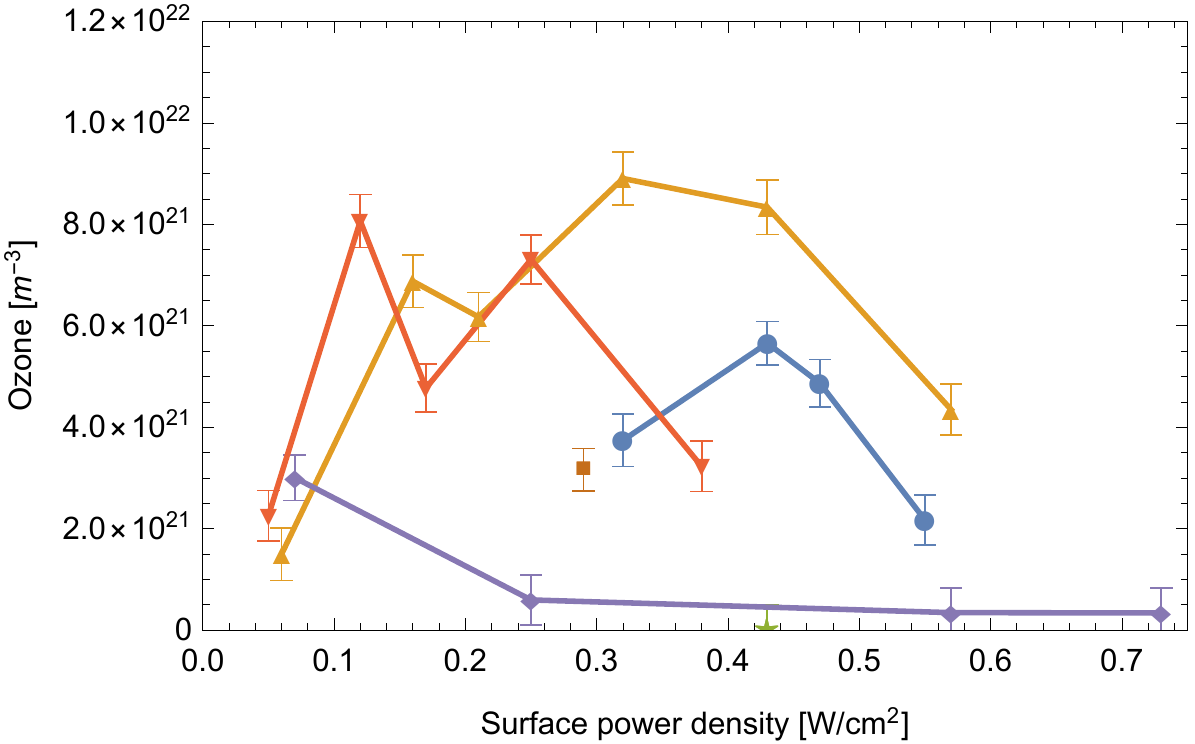}
      \begin{picture}(1,1)
	\centering
		\put(5,35){\includegraphics[width=1.2cm]{ODLeg.pdf}}
	\end{picture}
\end{minipage}
\caption{Ozone production dynamics of all plasma sources without an air flow. Saturation concentration (left) and concentration after \SI{15}{min} (right)}
\label{fig:15min}
\vspace*{4pt}
\hrulefill
\end{figure*}

There were three different dynamics of the ozone concentration. As figure \ref{fig:Trends} shows, the concentration is constant after reaching the saturation level. But when the plasma source heats itself up, there was a negative linear trend. In case of the large high voltage electrodes (5 \& 6) an exponential drop of the concentration occurs, which leads to the assumption, that the formation of RNS quenches the ozone production and therefore it could be an indication to the RNS production rate \cite{Shimizu2012,Li2019}.\\
The transition from ozone to nitrogen mode was expected to be constantly around \SI{0.1}{W/cm^2} \cite{Shimizu2012}, the measurements indicate, that there was a strong influence of the plasma source concept. On the one hand, (1) and (2) can reach \SI{0.32}{W/cm^2} without a drop of the ozone density. On the other hand, even low surface power densities of \SI{0.05}{W/cm^2} can lead to a decrease of the concentration in the case of geometry (5). After \SI{15}{min} higher power densities lower the concentration to a non-significant level for plasma source (3) and (5). This unveils that the transition mode had no constant power surface density value and depends on the geometric parameters of the plasma sources. Hence, the power density is not the only predictor of the ozone concentration.\\
The dynamics of RNS formation is based on the assumption, that their dynamics were the same as the dissociation rate of the ozone density. The quenching of the ROS with higher power is in line with previous findings \cite{Shimizu2012}.\\
Since the air flow was a form of an active cooling, the thermal effects were stronger without it and the curve is similar to the blue in figure \ref{fig:Trends}. And according to the results of the previous section, the destabilization of the plasma by the air flow leads to a lower ozone production after few minutes, because of the small number of filaments. To display the dynamics of the concentration in dependency of the power density and time, figure \ref{fig:15min} shows the saturation concentrations and the values after \SI{15}{min}.

\subsection{Inactivation Capability}
\begin{figure}[t!]
   \centering
   \includegraphics[width=0.45\textwidth]{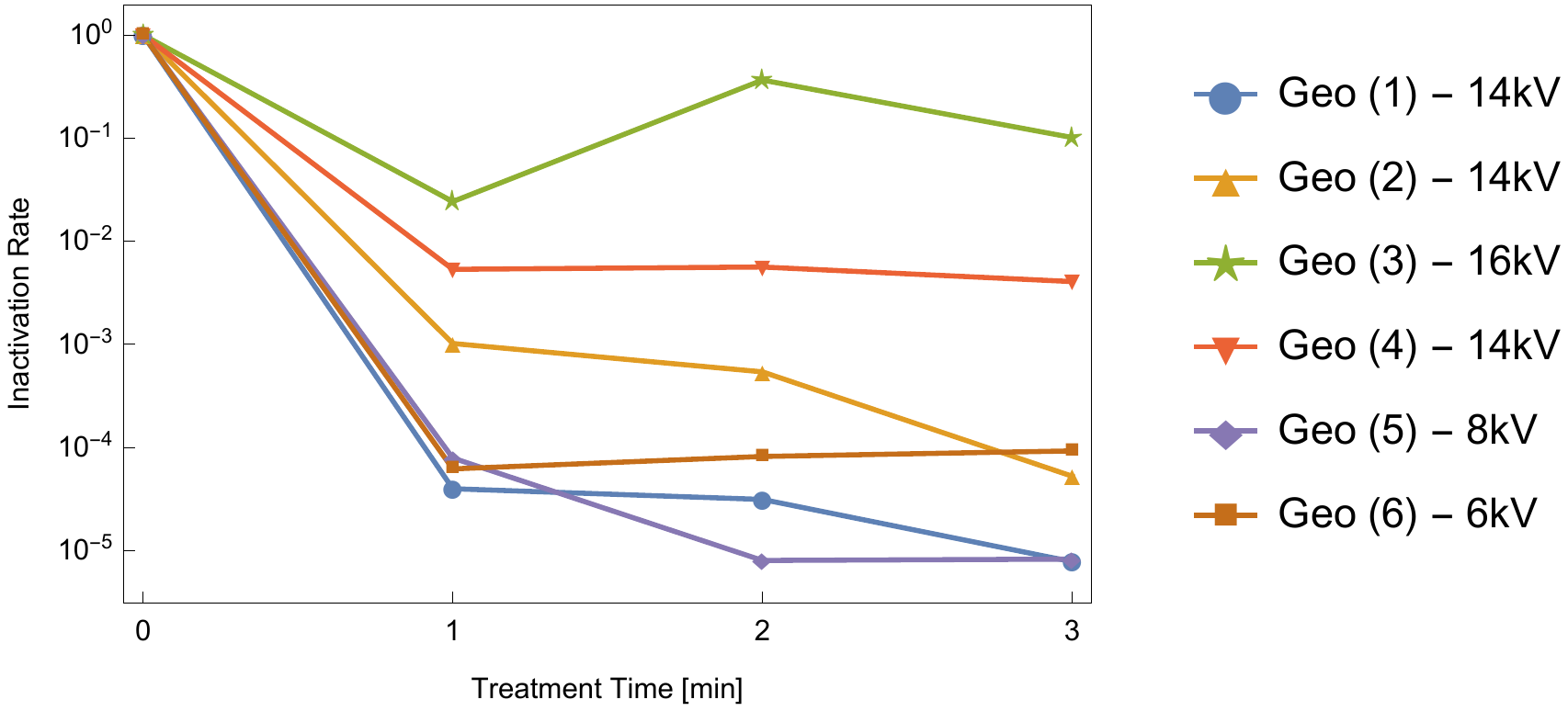}
\caption{Inactivation rates of all plasma sources, (3) and (4) with airflow}
\label{fig:steri}
\end{figure}

The biocidial effect is a desired capability of the RONS produced by the plasma. To understand the implications of the electrode geometry for the inactivation, a microbiological study was conducted to evaluate the differences of inactivation capability between the plasma sources. This shows which plasma source concept is the best. But a much wider theoretical contribution can be made by the correlation of these results to the other measurements.\\
Before the inactivation performance was estimated, the data were screened. Plates with too much CFU (\(\ge\)300) or outliers were removed from the statistics. Then for every test series the initial bacteria load was evaluated with the control plates. The counted CFUs divided by the initial CFUs is the inactivation performance, which was estimated for every agar plate.\\
Therefore, figure \ref{fig:steri} shows the means of the inactivation performance for every plasma source ((3) and (4) with the air flow) and exposure time.
Similar to the gas composition in figure \ref{fig:15min}, figure \ref{fig:steri} reveals differences of the inactivation capability, even when the power consumption was constant. Remarkable is the difference of the concepts (1), (5) to (2). Even though (2) has a higher ozone concentration, visible in figure \ref{fig:15min}, (1) inactivates up to \(10^4\) CFUs within a minute, which corresponds to a D-value of \SI{15}{s}. This capability is comparable to plasma source (5), which uses a lower surface power density but reaches the quenching effect. Those sources have inactivation rates of \(10^{-4}\) to \(10^{-5}\), which were the best results of all geometries. The RNS of (5) were expected to cause a higher inactivation performance \cite{Pavlovich_2014}. Contrarily, the inactivation rate does not improve from minute 2 to 3, when the quenching effect takes place.\\
The values, which represent decreasing inactivation rates with increasing exposition times, as (3) and (6) show, were not theoretical supported and were a result of statistical fluctuations.

\subsection{Air Flow Analysis}
All data are analyzed with respect to the air flow (\SI{1}{m^3/h}). The electrical parameters are neither changed by the slits nor by an applied air flow through those. A total different behavior unveil the results of the time resolved ignition behavior. The slits bring up virtual electrodes and the air flow destabilizes the formation of filaments within the slits. This is visible in figure \ref{fig:Geo3Airflow}. During the first \SI{10}{s} after the air flow was applied, the plasma becomes weaker and does not ignite homogeneously. The reason can be either a deformation of the filaments or a higher recombination due to the higher pressure by the air flow. The air flow changes the formation of filaments and the ozone production. Therefore, the microbiological studies reflect those results by the inactivation capability, which was reduced three to four orders of magnitude. The suggested advantages of this (high) air flow compared to the diffusion has to be neglected.

\section{Conclusion}
This work verifies the applicability of thin PCBs as plasma sources. The resulting advantages are the fast prototyping compared to dielectrics like glass or ceramics and the lower power consumption compared to volume DBDs \cite{CEPLANT2002}. Additionally, PCBs have laminated electrodes, therefore it is not possible that gases can be between the electrodes and the dielectric.\\
The capacity is given by the amount of charges, which can accumulate above the high voltage electrode.
Increasing voltages lead to a higher power consumption and higher capacities, because of the higher electron density. Surface power densities of \SI{0.05}{W/cm^2} to \SI{0.57}{W/cm^2} were achievable. 
The ignition behavior depends on the distance between the electrodes. Longer discharges lead to arc-like filaments and due to the lower electric field, to a shorter duration of the plasma ignition. All plasma sources have in common, that the discharge at the beginning of the negative half-cycle was the strongest and consists of simultaneously igniting filaments. Additionally, this analysis verifies the concept of the virtual electrodes, which lead to a discharge above the dielectric and not in the slits at low voltages. A higher electric field leads to accumulation of electrons at the edges of and filaments over the slits.\\
Geometry (2) produces the highest ozone concentration of \SI{435}{ppm}. The ozone density shows as well as the ozone production a positive dependency on the supply voltage. The quenching of ozone occurs with concept (5/6) and is an indication of the RNS production. The formation of curved filaments occurs only with this concept. The linkage of the filaments' shape to the gas-composition will become addressed by further research.\\
The microbiological study unveils inactivation rates of \(10^{-4}\) to \(10^{-5}\) as the best results of all geometries. Therefore, D-values of \SI{15}{s} were achievable. The air flow in this case destabilizes the plasma and extinguishes the filaments. Due to that, the worst case (3) show a drop of the inactivation performance of three to four orders of magnitude. Therefore, it can be concluded that the plasma chemistry and the inactivation capability is influenced by the electrode geometry and the surface power density as well. The air flow does not improve the plasma technology until now.\\
Finally, it is to state that the electrode geometry has to be defined with respect to the scope of application. The results recommend displaced electrodes to improve the inactivation capability. Yet, ozone formation was increased, when there are parallel electrodes and spaces with a darker plasma glow.\\
The PCB-technology is a promising process for plasma source development. Yet, epoxy resin based dielectrics have a limited lifetime of \SI{15}{h}. Therefore further development effort will be invested in PCB based SMDs with other dielectric materials. This will make the plasma inactivation the next disinfection technology of the future, which will reduce nosocomial infections and durations of reprocessing cycles.
\begin{table}[t!]
  \centering
    \caption{Overview of the four evaluations}
  \begin{tabular}{c||cccc}
  \bfseries Geo. & \bfseries Power  & \bfseries Filaments' & \bfseries Ozone & \bfseries Log-\\
    \bfseries & \bfseries density [W/cm\(\mathbf{^2}\)] & \bfseries shape & \bfseries [ppm] & \bfseries reduction\\\hline\hline\\
  (1) &0.32\,-\,0.55& arc-like & 212 & 5\\
  (2) &0.06\,-\,0.57& short & 435 & 4\\
  (3) &0.43& virtual electrodes & 376 & 1\\
  (4) &0.05\,-\,0.45& virtual electrodes & 125 & 2 \\
  (5) &0.07\,-\,0.57& curved & 82 & 5\\
  (6) &0.29& curved & 184 & 4\\
\end{tabular} 
\label{tab:Zus}
\end{table}
\bibliographystyle{IEEEtran.bst}
\bibliography{References.bib}
\end{document}